\documentclass[%
reprint,
amsmath,amssymb, 
aps,
prl,
]{revtex4-2}
\usepackage{graphicx, color}
\usepackage{dcolumn}
\usepackage{bm}
\usepackage{appendix,float}
\usepackage[colorlinks,urlcolor=blue,linkcolor=blue,anchorcolor=blue,citecolor=blue]{hyperref} 

\begin{document}
\preprint{APS/123-QED}
\title{Anomalous charge density wave in altermagnetism}
\author{Zi-Hao Ding$^{1,2}$}
\author{Lei Wang$^{3}$}
\author{Zhen-Feng Ouyang$^{1,2}$}
\author{Jingsi Qiao$^{4}$}
\author{Ze-Feng Gao$^{1,2}$}
\author{Wei Ji$^{1,2}$}
\author{Kai Liu$^{1,2}$}
\author{Peng-Jie Guo$^{1,2}$}
\email{guopengjie@ruc.edu.cn}
\author{Zhong-Yi Lu$^{1,2,5}$}
\email{zlu@ruc.edu.cn}
\affiliation{1. School of Physics and Beijing Key Laboratory of Opto-electronic Functional Materials $\&$ Micro-nano Devices, Renmin University of China, Beijing 100872, China}
\affiliation{2. Key Laboratory of Quantum State Construction and Manipulation (Ministry of Education), Renmin University of China, Beijing 100872, China}
\affiliation{3. School of Physical Science and Technology, Inner Mongolia University, Hohhot 010021, China}
\affiliation{4. School of Integrated Circuits and Electronics $\&$ Advanced Research Institute of Multidisciplinary Sciences, Beijing Institute of Technology, Beijing 100081, China}
\affiliation{5. Hefei National Laboratory, Hefei 230088, China}
\date{\today}
\begin{abstract}

Exploring the intricate interplay between magnetism and charge density waves has long been a fundamental pursuit at the forefront of condensed matter research.
In this letter, based on symmetry analysis and first-principles calculations, we propose for the first time that anomalous charge density wave can be realized in two-dimensional altermagnetic WO. The anomalous charge density wave is characterized by three key features: (i) Unlike conventional charge density wave, whose stabilization is driven by the opening of a gap near the Fermi level, the anomalous charge density wave is stabilized by the occupied states with energies shifting lower far away from the Fermi level; (ii) the anomalous charge density wave increases the density of states near the Fermi level and then enhances—rather than diminishes—the metallicity of materials; (iii) altermagnetism plays a crucial role in stabilizing anomalous charge density wave. Thus, our work offers a pathway for exploring both the realization and the underlying mechanisms of anomalous charge density waves in magnetic systems.

\end{abstract}

\maketitle

{\it Introduction.} Altermagnetism, a new magnetic phase distinct from both ferromagnetism and antiferromagnetism, has been proposed theoretically and confirmed experimentally \cite{PRX-1,PRX-2,MnTe-1,MnTe-2,MnTe-3,CrSb-1,CrSb-2,CrSb-3,CrSb-4,KV2Se2O}. Altermagnetic materials exhibit momentum-space anisotropic spin splitting without spin–orbit coupling (SOC) and real-space antiparallel spin alignment with zero net total magnetic moment, thereby sharing the advantages of both ferromagnetic and antiferromagnetic materials. Consequently, numerous exotic phenomena can emerge in altermagnetic materials, such as anisotropic spin splitting leading to spin currents \cite{SST-PRL2021,SST-NE2022,SST-PRL2022,SST-PRL2022-2}, giant and tunneling magnetoresistance \cite{GMR-TMR,TMR-1,TMR-2}, and piezomagnetic effects \cite{piezomagnetism-NC}. With SOC, due to the break of the time-reversal symmetry, altermagnetic materials can realize the anomalous Hall \cite{sciadv.aaz8809,vsmejkal2022anomalous,npj,AHE-hou2023}, anomalous magneto-optical \cite{magneto-optica-1,magneto-optica-2}, anomalous Nernst \cite{Nernst}, anomalous thermal Hall \cite{CTT} and crystal valley Hall \cite{CVHE-tan} effects. On the other hand, the combination of altermagnetism with other matter phases has also attracted widespread interest. The anisotropic spin-splitting altermagnetism without time-reversal symmetry merging with topology gives rise to new topological phases\cite{npj,MirrorChern, QCVHE-tan, NSR, higher-order, Type-II-QSHI,QSHC-YAO}. Coupling time-reversal-broken altermagnetism with superconductivity can lead to nontrivial superconducting states \cite{superconductivity-1,superconductivity-2}. Altermagnetic ferroelectricity enables electrical control of magnetism \cite{Ferroelectric-1,Ferroelectric-2,Ferroelectric-3,Ferroelectric-4}. 

\begin{figure}[htbp]
 \centering
 \includegraphics[width=8.5cm]{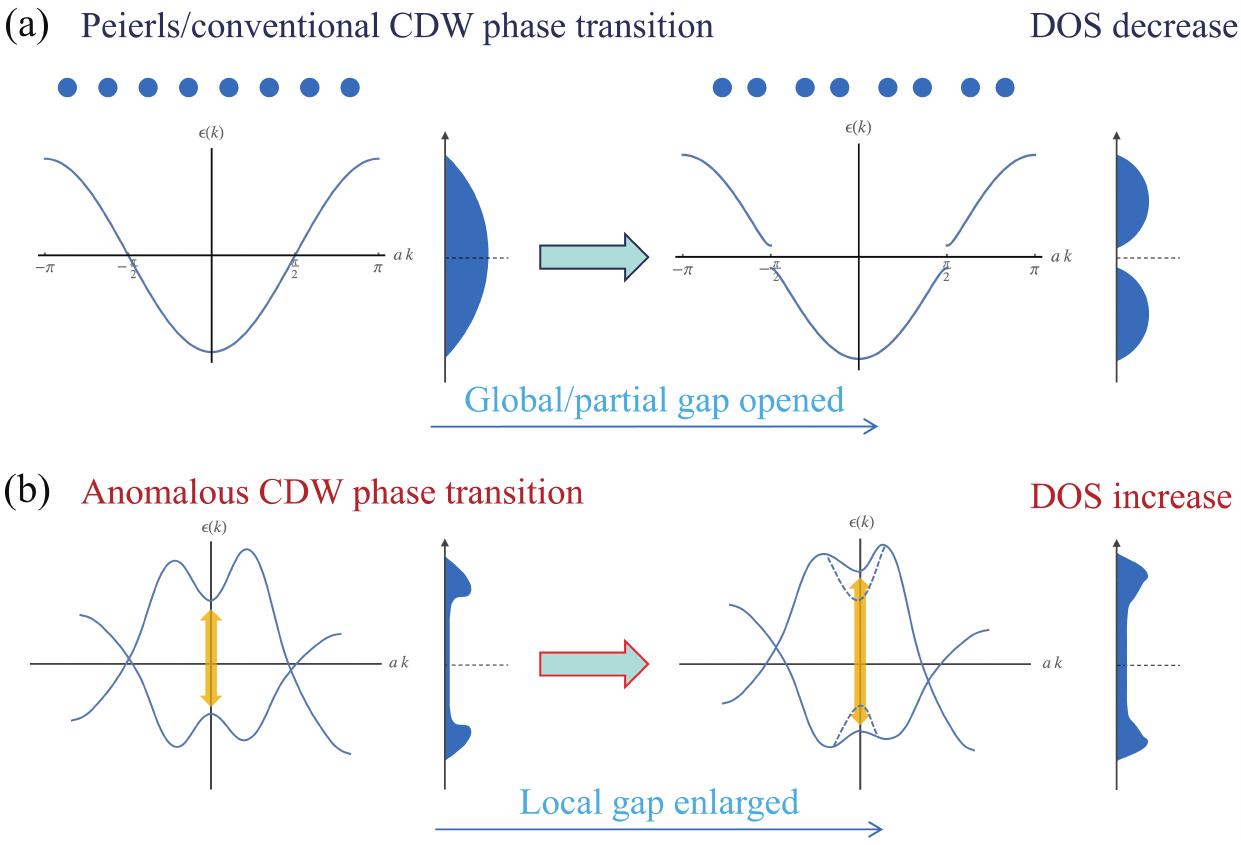}
 \caption{Schematic diagram of the (a) Peierls/conventional CDW phase transition and  (b) Anomalous CDW phase transition, showing electronic band structures and density of states (DOS) before and after the CDW transition.}
 \label{fig1}
 \end{figure}

Charge density wave (CDW) is a common quantum phase that accompanies lattice distortion and generally drives a metal-to-insulator transition or a reduction in the density of states (DOS) near the Fermi level (i.e., poorer metallicity) \cite{CDW-BOOK,CDW-1,CDW-Classification,CDW-misconceptions,CDW-FSN}, as illustrated in Fig.~\ref{fig1}{(a)}. Due to strong electron-phonon coupling, the suppression of the CDW often precedes the emergence of superconductivity \cite{CDW-SC1,CDW-SC2}. An intriguing possibility arises if, instead of weakening metallicity, the CDW actually enhances metallicity, thereby transforming the traditionally competitive relationship between CDW and superconductivity into a cooperative one. Conventionally, CDW stabilization stems from the opening of a gap near the Fermi level, which lowers the total energy at the cost of reduced metallicity. If the stability of CDW phase is instead driven by the downward shift of locally occupied bands far from the Fermi level, the metallicity of materials may be enhanced, as illustrated in Fig.~\ref{fig1}{(b)}. We term such a CDW as an anomalous CDW. In nonmagnetic materials, CDWs typically degrade metallicity; the introduction of magnetism may facilitate the realization of anomalous CDWs. Very recently, the combination of altermagnetism and CDWs has begun to be explored \cite{CsCr3Sb5,CoNb4Se8,magnetic-field-tunable}. Thus, whether or not is the anomalous CDW realized in an altermagnetic material? 



In this letter, based on symmetry analysis and first-principles calculations, we propose for the first time that anomalous CDW can be realized in two-dimensional altermagentic WO. Moreover, our calculations reveal that altermagnetism plays a crucial role in stabilizing the $\sqrt{2}\times\sqrt{2}$ CDW in monolayer WO. Thus, our work offers a pathway for exploring both the realization and the underlying mechanisms of anomalous charge density waves in magnetic systems.

{\it Results and analysis.}
Recently, monolayer CrO and MoO have been predicted to be bipolarized Weyl semimetals with altermagnetism, and they will transition to the Chern insulator phase under strain \cite{CrO,npj,MoO,QCVHE-tan}. Monolayer CrO is a perfect bipolarized Weyl semimetal, while the overlap of the valence and conduction bands in monolayer MoO leads to the presence of both electron and hole Fermi surfaces. Therefore, monolayer MoO exhibits better metallic properties than monolayer CrO. If W atoms substitute Mo atoms, monolayer WO should have stronger metallic properties. Additionally, 5$d$ transition metal oxides may lead to structural distortions \cite{khomskii2014transition}, so monolayer WO provides the possibility for the coexistence of altermagnetism and anomalous CDW. 

\begin{figure}[htbp]
\centering
\includegraphics[width=8.5cm]{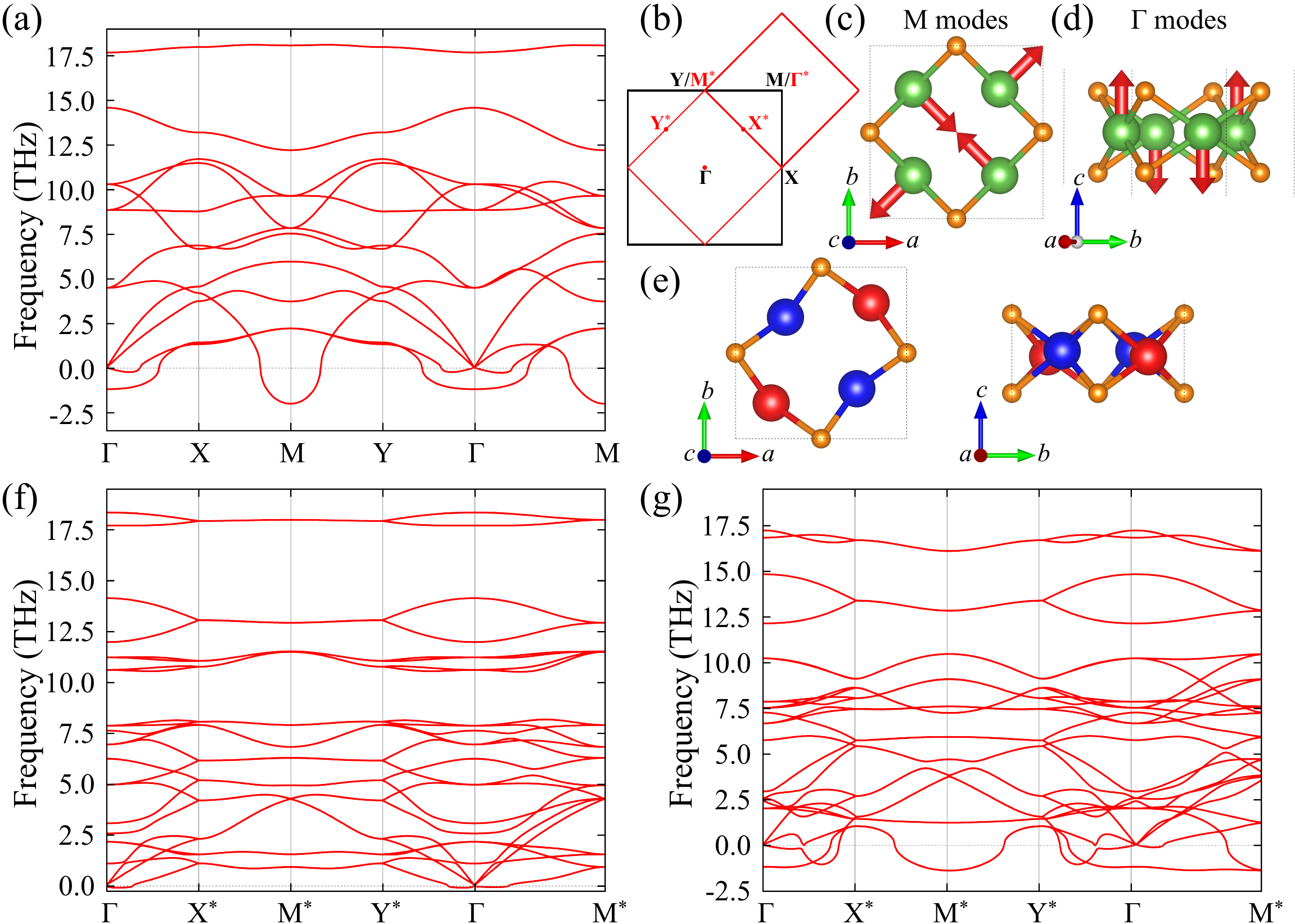}
\caption{(a) Phonon spectrum for the normal phase of monolayer WO. (b) First Brillouin zone (BZ) of monolayer WO for primitive cell (black solid lines) and $\sqrt{2}\times\sqrt{2}$ supercell (red solid line). (c) and (d) are the imaginary vibration modes at the M and $\Gamma$ points, respectively. (e) Top and side views for the CDW phase of monolayer WO with altermagnetism. (f) and (g) are phonon spectra of the CDW phase for altermagnetic and nonmagnetic monolayer WO, respectively.}
\label{fig2}
\end{figure}

Similar to the monolayer CrO, the crystal structure of monolayer WO has the $P4/mmm$ space group symmetry, corresponding to the $D_{4h}$ point group. The primitive cell of monolayer WO contains two W atoms and two O atoms, with an optimized lattice constant of 3.696 \text{\AA}. Moreover, the detailed calculations reveal that the magnetic ground state of monolayer WO remains $d$-wave altermagnetic across a range of Hubbard U, consistent with monolayer CrO (the results and analyses are provided in the Supplementary Material (SM) \cite{SM}). Then, we investigate the dynamical stability of monolayer WO based on altermagnetic structure. We calculate the phonon spectrum of WO, as shown in Fig.~\ref{fig2}{(a)}. From Fig.~\ref{fig2}{(a)}, WO has two phonon modes with imaginary frequencies respectively at the high-symmetry M and $\Gamma$ points, and the imaginary frequency at the M point is larger than that at the $\Gamma$ point. We speculate that the imaginary frequencies might be caused by a $\sqrt{2}\times\sqrt{2}$ CDW. To verify this speculation, we first calculate the phonon spectrum of the $\sqrt{2}\times\sqrt{2}$ supercell. Under the $\sqrt{2}\times\sqrt{2}$ supercell, the M point of the primitive cell Brillouin zone (BZ) is folded to the $\Gamma$ point, as shown in Fig.~\ref{fig2}{(b)}. Therefore, the vibration mode with imaginary frequency at the M point of the primitive cell BZ is folded to the $\Gamma$ point. Then, we plot these two vibration modes with imaginary frequency, as shown in Fig.~\ref{fig2}{(c)} and \ref{fig2}{(d)}, respectively. From Fig.~\ref{fig2}{(c)} and \ref{fig2}{(d)}, these two vibration modes originate from the vibrations of W atoms. The vibration mode with imaginary frequency at the $\Gamma$ point is the relative vibration of W atoms along the $z$ direction, and the vibration mode with imaginary frequency at the M point is the in-plane relative vibration of W atoms. Based on these two vibration modes, we obtain the crystal structure with lattice distortion through structural optimization, as shown in Fig.~\ref{fig2}{(e)}.

\begin{figure*}[htbp]
\centering
\includegraphics[width=16 cm]{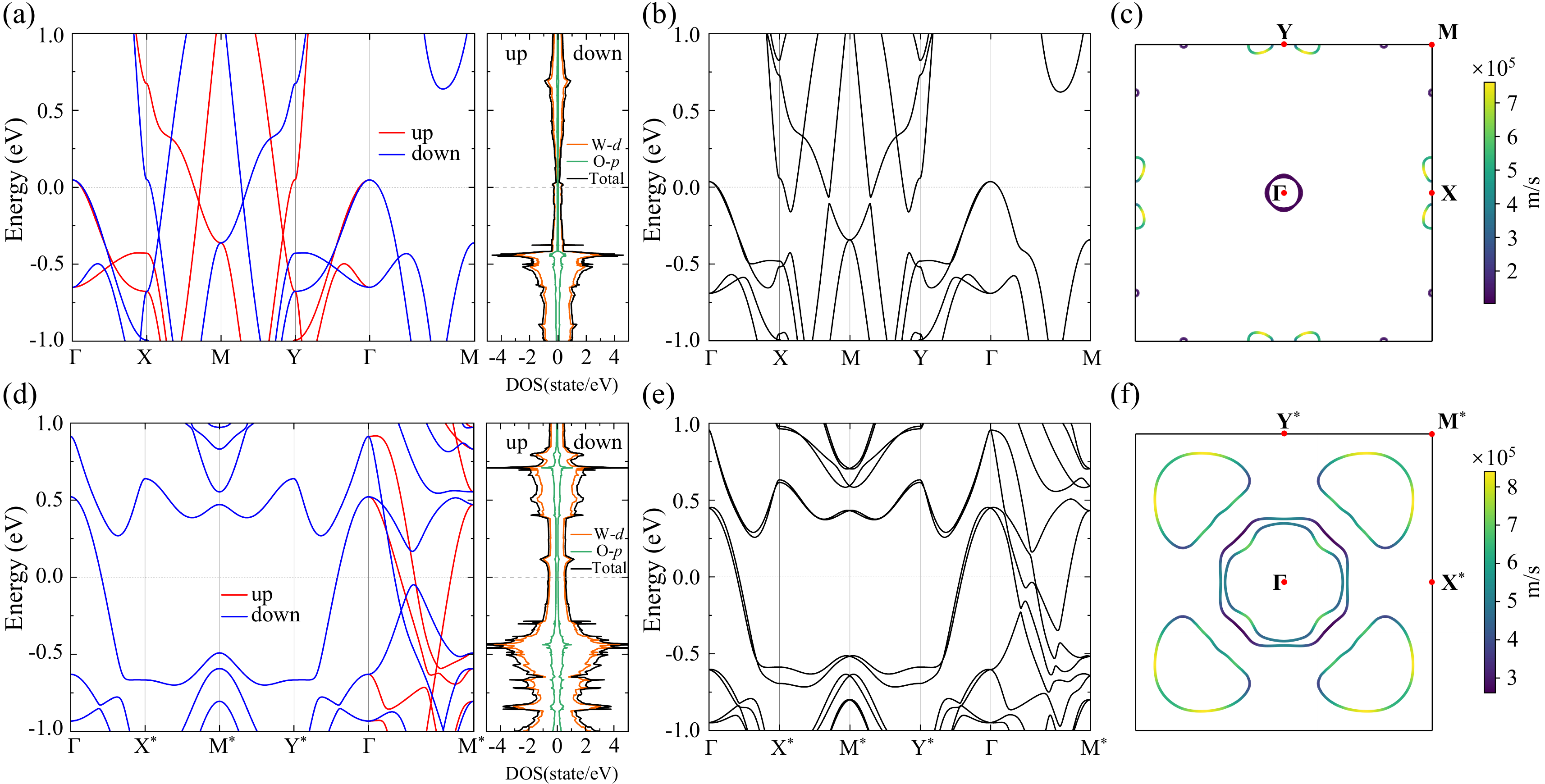}
\caption{(a) and (d) are electronic band structures and density of states without SOC for normal phase and CDW phase of monolayer WO, respectively. the red and blue lines represent spin-up and spin-down bands, respectively. (b) and (e) are electronic band structures with SOC for normal phase and CDW phase of monolayer WO, respectively. (c) and (f) are Fermi surfaces with SOC for normal phase and CDW phase of monolayer WO, respectively. The different colors represent the magnitudes of the carrier group velocities.}
\label{fig3}
\end{figure*}

After lattice distortion, the symmetry of crystal structure changes from the high-symmetry $P4/mmm$ to the low-symmetry $P$-$42_{1}m$ space group symmetry, corresponding to the $D_{2d}$ point group. Interestingly, the magnetic ground state of monolayer WO with lattice distortion remains $d$-wave altermagnetic (Fig.~\ref{fig2}{(e)}) (the results and analyses are provided in the SM \cite{SM}). Then, we calculate the phonon spectrum for this $\sqrt{2}\times\sqrt{2}$ distorted supercell with altermagnetism, as shown in Fig.~\ref{fig2}{(f)}. From Fig.~\ref{fig2}{(f)}, there are no imaginary frequencies in the phonon spectrum. Note: The tiny imaginary frequency at the $\Gamma$ point is caused by the finite displacement calculation method, and the density-functional perturbation theory based on the Quantum Espresso \cite{giannozzi2009} can completely eliminate the imaginary frequency at $\Gamma$ point (the results and analyses are provided in the SM \cite{SM}). Therefore, the two imaginary frequencies at the M and $\Gamma$ points in the normal phase of monolayer WO are indeed caused by a $\sqrt{2}\times\sqrt{2}$ CDW. That is to say, monolayer WO is a material in which $\sqrt{2}\times\sqrt{2}$ CDW and altermagnetism coexist. On the other hand, we also calculated the phonon spectrum of nonmagnetic monolayer WO in the CDW phase, as shown in Fig.~\ref{fig2}{(g)}. From Fig.~\ref{fig2}{(g)}, there are pronounced imaginary frequencies, reflecting that nonmagnetic monolayer WO does not host a $\sqrt{2}\times\sqrt{2}$ CDW. Therefore, altermagnetism plays a crucial role in stabilizing the $\sqrt{2}\times\sqrt{2}$ CDW in monolayer WO.

A natural question is what causes the $\sqrt{2}\times\sqrt{2}$ CDW in altermagnetic WO. This requires us to further investigate the electronic structure of altermagnetic WO. In the subsequent electronic structure calculations, we include a Hubbard U = 2.2 eV to account for the on-site Coulomb repulsion among the 5$d$ orbitals of W atoms according to previous studies \cite{U1,U2}. We first calculate the electronic band structure and density of states of altermagnetic WO without SOC in the normal phase, as shown in Fig.~\ref{fig3}{(a)}. From Fig.~\ref{fig3}{(a)}, monolayer WO is an altermagnetic semimetal with four pairs of crossing points near the Fermi level. Two pairs of crossing points with opposite polarization are protected by the spin symmetry $\left\{ E|| C_{2y} \right\}$ along the X-M direction, and the other two pairs of crossing points with opposite polarization are protected by the spin symmetry $\left\{ E|| C_{2x} \right\}$ along the Y-M direction. Therefore, monolayer WO is a bipolarized Weyl semimetal, similar to monolayer Fe$_2$WTe$_4$ \cite{QCVHE-tan}. On the other hand, since the bands near the Fermi level are mainly contributed by the 5$d$ orbitals of W atoms (Fig.~\ref{fig3}{(a)}), monolayer WO likely exhibits strong SOC effect.


\begin{figure}[htbp]
\centering
\includegraphics[width=8.5cm]{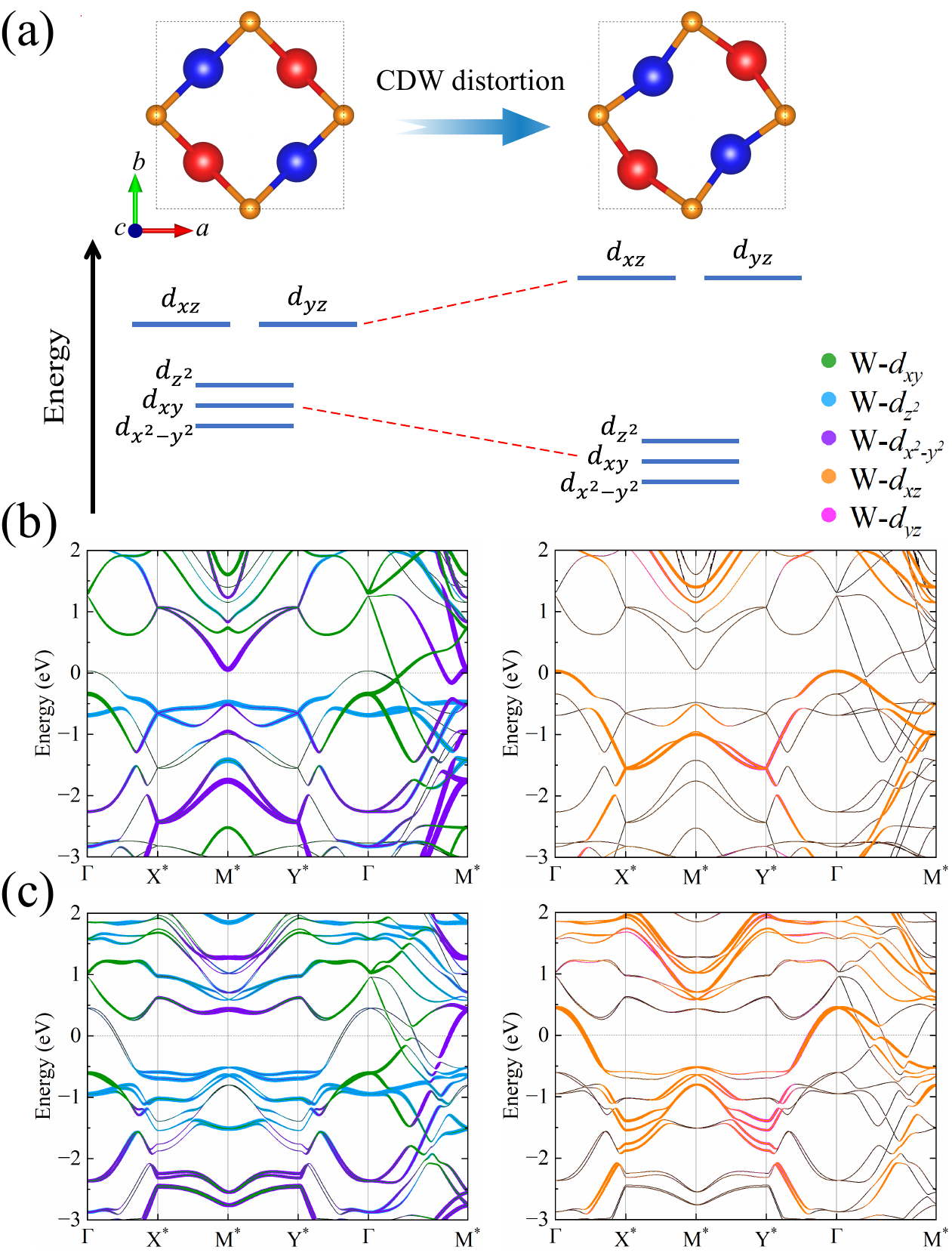}
	\caption{(a) Schematic diagram of crystal field splitting in 5$d$ Orbitals of W atoms. The $d$-orbital projected band structures with SOC for (b) the $\sqrt{2}\times\sqrt{2}$ normal phase and (c) the CDW phase of monolayer WO.}
	\label{fig4}
\end{figure}
 
Before calculating the band structure with SOC, we need to first determine the direction of the magnetic easy axis of monolayer WO. This is because the direction of the magnetic easy axis determines the magnetic point group symmetry, which in turn has a significant impact on the electronic structure and topological properties of monolayer WO. Our calculations show that the magnetic easy axis of WO is along the $z$ direction. So altermagnetic WO breaks the $C_{2x}$ and $C_{2y}$ symmetries under SOC, which in turn leads to the Weyl points near the Fermi level opening gap. Next, we calculate the electronic structure of monolayer WO with SOC, as shown in Fig.~\ref{fig3}{(b)}. Comparing Fig.~\ref{fig3}{(a)} and \ref{fig3}{(b)}, monolayer WO indeed exhibits strong SOC effect, especially the two pairs of Weyl points near the X and Y points open a band gap of 266 meV. However, monolayer WO remains a semimetal, with its corresponding Fermi surface shown in Fig.~\ref{fig3}{(c)}. From Fig.~\ref{fig3}{(c)}, the hole Fermi surface is near the $\Gamma$ point, while the electron Fermi surfaces are located near the X and Y points. Clearly, the hole Fermi surface and the electron Fermi surface cannot be nested by the ($\pi$, $\pi$) wave vector, but the electron Fermi surfaces near the X and Y points can be nested by the ($\pi$, $\pi$) wave vector. However, calculation of the Lindhard susceptibility function shows that monolayer WO exhibits no $\sqrt{2}\times\sqrt{2}$ CDW instability driven by Fermi surface nesting (the results and analyses are provided in the SM \cite{SM}). In view of the strong phonon softening observed in the normal phase, the $\sqrt{2}\times\sqrt{2}$ CDW in monolayer WO likely arises from the synergy of electron-phonon coupling with altermagnetism.



Is the CDW in monolayer WO an anomalous CDW? To verify it, we calculate the electronic structure of the altermagnetic WO without SOC in the CDW phase, as shown in Fig.~\ref{fig3}{(d)}. Unlike the normal phase, the bands along the $\Gamma$-X and $\Gamma$-Y directions in the CDW phase of WO are spin degenerate, while the bands along the $\Gamma$-M direction are spin split (Fig.~\ref{fig3}{(d)}). This is because the spin symmetries $\left\{ C_2|| C_{2x} (1/2, 1/2) \right\}$  and $\left\{ C_2|| C_{2y} (1/2, 1/2) \right\}$ protect the spin degeneracy of the bands along the $\Gamma$-X and $\Gamma$-Y directions in the CDW phase of monolayer WO. On the other hand, since the states near the Fermi level are still contributed by the 5$d$ orbitals of W atoms, we also calculate the electronic band structure of the CDW phase of WO with SOC (Fig.~\ref{fig3}{(e)}). Although the SOC splits the crossing points along the $\Gamma$-M direction and the spin degeneracy on the high-symmetry lines, the splitting is not significant. Importantly, the DOS in the CDW phase is 2.32 times that of the normal phase at the Fermi level, transforming monolayer WO from a semimetal into a metal. Moreover, we present the Fermi surface of the CDW phase of WO, which intuitively shows that the CDW phase of WO becomes a good metal (Fig.~\ref{fig3}{(f)}). Therefore, the anomalous CDW can be realized in altermagnetic WO. In addition, we also calculate the electronic structure of nonmagnetic monolayer WO in the CDW phase (the results and analyses are provided in the SM \cite{SM}). In the nonmagnetic phase, monolayer WO remains metallic; its bands differ markedly from those in the altermagnetic phase, yet these states near the Fermi level are still dominated by the 5$d$ orbitals of W atoms. Despite these similarities, the $\sqrt{2}\times\sqrt{2}$ CDW is unstable in the nonmagnetic case. This further underscores the crucial role played by altermagnetism in enabling the anomalous CDW.

Another important question is why the CDW in altermagnetic WO is anomalous. To answer this question, we first compare the crystal-field splittings of the 5$d$ orbitals from W atoms before and after the lattice distortion. In the $\sqrt{2}\times\sqrt{2}$ normal-phase monolayer WO, W sits in a rectangular crystal field whose $d_{xz}$ and $d_{yz}$ orbitals lie at higher energy, while $d_{x^2-y^2}$, $d_{xy}$, and $d_{z^2}$ are lower (Fig.~\ref{fig4}{(a)}). After the CDW distortion, W sits in a distorted rectangular field that pushes $d_{xz}$ and $d_{yz}$ even higher and pulls $d_{x^2-y^2}$, $d_{xy}$, and $d_{z^2}$ to lower energies (Fig.~\ref{fig4}{(a)}). Moreover, the large downward shift stabilizes the CDW phase. Then, we plot band structures with $d$-orbital projection for normal-phase and CDW-phase WO, as shown in Fig.~\ref{fig4}{(b)} and \ref{fig4}{(c)}, respectively. They confirm the higher-lying $d_{xz}$ and $d_{yz}$ and lower-lying $d_{x^2-y^2}$, $d_{xy}$, and $d_{z^2}$ levels, and show that after the CDW transition $d_{xz}$ and $d_{yz}$ rise further while $d_{x^2-y^2}$, $d_{xy}$, and $d_{z^2}$ down further (Fig.~\ref{fig4}{(b)} and \ref{fig4}{(c)}). Notably, the bands crossing the Fermi level are dominated by the $d_{xz}$ and $d_{yz}$ orbitals (Fig.~\ref{fig4}{(b)} and \ref{fig4}{(c)}); their upward shift in the CDW phase enlarges the Fermi surface and turns WO into a metal. Therefore, the origin of the anomalous CDW in monolayer WO is fully consistent with the proposed mechanism illustrated in the schematic (Fig.~\ref{fig1}{(b)}).

According to the results and analysis above, the anomalous CDW in monolayer WO exhibits three distinctive features: (i) Altermagnetism plays a crucial role in stabilizing anomalous CDW; (ii) Unlike conventional CDW, whose stabilization is driven by the opening of a gap near the Fermi level, the anomalous CDW is stabilized by the occupied states with energies shifting lower far away from the Fermi level; and (iii) The second feature increases the density of states near the Fermi level, thereby driving the anomalous semimetal-to-metal transition in monolayer WO induced by CDW. Furthermore, the three distinctive features of the anomalous CDW in monolayer WO are not confined to altermagnetic materials; they may also extend to other magnetic phases, such as collinear antiferromagnetism and unconventional non-collinear antiferromagnetism. Our work therefore offers a promising avenue for exploring anomalous charge density waves in magnetic systems more broadly.



\begin{acknowledgments}
We thank Jun-Wei Liu for valuable discussions. This work was financially supported by the National Natural Science Foundation of China (Grant No.12434009, No.12204533, No.12304165 and No.62476278), the National Key R$\&$D Program of China (Grant No. 2024YFA1408601), the Fundamental Research Funds for the Central Universities, and the Research Funds of Renmin University of China (Grant No. 24XNKJ\textsubscript{1}5). Computational resources have been provided by the Physical Laboratory of High Performance Computing at Renmin University of China.

Z.-H. D. and L. W. contributed equally to this work.
\end{acknowledgments}

\nocite{*}

\bibliography{Reference}

\end{document}